\documentclass[aps,prl,twocolumn,superscriptaddress,showpacs]{revtex4}

\usepackage{graphicx}
\usepackage{afterpage}

\begin{document}


\title{Anisotropic superconductivity in layered LaSb$_2$: the role
of phase fluctuations}



\author{S. Guo}
\affiliation{Department of Physics and Astronomy, Louisiana State
University, Baton Rouge, Louisiana 70803, USA}

\author{D.P. Young}
\affiliation{Department of Physics and Astronomy, Louisiana State
University, Baton Rouge, Louisiana 70803, USA}

\author{P. W. Adams}
\affiliation{Department of Physics and Astronomy, Louisiana State
University, Baton Rouge, Louisiana 70803, USA}

\author{X. S. Wu}
\affiliation{Department of Physics and Astronomy, Louisiana State
University, Baton Rouge, Louisiana 70803, USA}

\author{Julia Y. Chan}
\affiliation{Department of Chemistry, Louisiana State University,
Baton Rouge, Louisiana 70803, USA}

\author{G.T. McCandless}
\affiliation{Department of Chemistry, Louisiana State University,
Baton Rouge, Louisiana 70803, USA}

\author{J.F. DiTusa}
\email[]{ditusa@phys.lsu.edu}
\affiliation{Department of Physics and Astronomy, Louisiana State
University, Baton Rouge, Louisiana 70803, USA}


\date{\today}

\begin{abstract}
We present electrical transport and magnetic susceptibility
measurements of the highly anisotropic compound LaSb$_2$ observing a
very broad transition into a clean, consistent with type-I,
superconducting state with distinct features of 2
dimensionality. Application of hydrostatic pressure induces a 2- to
3-dimensional crossover evidenced by a reduced anisotropy and
transition width. The superconducting transition appears phase
fluctuation limited at ambient pressure with fluctuations observed for
temperatures greater than 8 times the superconducting critical
temperature.
\end{abstract}


\maketitle

Superconductivity in reduced dimensions has intrigued condensed matter
physicists for over 40 years. Highly anisotropic materials with
superconducting (SC) phases, such as TaS$_2$ and
NbSe$_2$\cite{gamble,prober,nagata,jerome}, as well as thin SC
metallic films{\cite{aoi,goldman,zhang} and organic
compounds\cite{singleton} were investigated to search for novel
properties stemming from dimensionality effects. More recent
discoveries of unconventional superconductivity in layered
cuprates\cite{bednorz}, MgB$_2$\cite{nagamatsu}, and iron
pnictides\cite{kamihara}, all possessing anisotropic crystal
structures, has highlighted the importance of the layered structure
and phase fluctuations\cite{emery,orenstein} in determining the SC and
normal properties of these compounds. Here we present resistivity,
magnetization, and AC susceptibility measurements on the highly
layered SC compound LaSb$_2$\cite{budko,lasb2dhva,ditusaoptical} which
has been of interest because of its magnetoresistive
properties\cite{younglasb2}.  Previous investigations of LaSb$_2$ show
no indication of competing order such as a charge density wave
transition\cite{ditusaoptical}. We present evidence that the ambient
pressure SC phase, in which only a minority of crystals display a
complete Meissner effect at low temperature, $T$, is characteristic of
poorly coupled two dimensional (2D) SC planes. The anisotropy is
reduced and the transition is dramatically sharpened as pressure is
applied indicating a crossover from a 2D to a more traditional 3D SC
phase. Our data demonstrate that the extraordinarily wide, and many
times incomplete, SC transition at ambient pressure likely results
from 2D phase fluctuations which persist for temperatures much lower
than the onset temperature for superconductivity, $T_{onset}$, that is
at $T$'s an order of magnitude larger than the global SC critical
temperature, $T_c$. This places LaSb$_2$ among a handful of
systems\cite{zhang,orenstein}} exhibiting phase fluctuation limited
superconductivity and is unusual in that it displays behavior
consistent with clean, type I, superconductivity\cite{yonezawa}.

LaSb$_2$ is a member of the RSb$_2$ (R=La-Nd, Sm) family of compounds
that all form in the orthorhombic, highly layered, SmSb$_2$
structure\cite{budko,younglasb2,sato} in which alternating La/Sb
layers and 2D rectangular sheets of Sb atoms are stacked along the
c-axis. These structural characteristics give rise to the anisotropic
physical properties observed in all the compounds in the RSb$_2$
series\cite{budko,younglasb2,lasb2dhva}. A large number of single
crystals of LaSb$_2$ grown from high purity La and Sb by the metallic
flux method were large flat, micaceous, plates, which are malleable
and easily cleaved.  Polycrystalline samples grown in crucibles using
a stoichiometric mixture of the constituents had $T_{onset}$
essentially identical to the crystals.  The SmSb$_2$ structure-type
with lattice constants of $a=0.6219$, $b=0.6278$, and $c=1.846$ nm,
was confirmed by single crystal X-ray diffraction. Resistivity,
$\rho$, measurements were performed with currents either in the
ab-plane or along the c-axis using standard 4-probe ac techniques at
17 or 27 Hz from $0.05\le T \le 300$ K.  Data presented here are from
single crystal samples with residual resistance ratios of 70-90
between 300 and 4 K. Magnetization, $M$, and susceptibility, $\chi$,
were measured with a commercial SQUID magnetometer for $T>1.75$ K and
a dilution refrigerator ac $\chi$ probe for $T \ge 50$ mK and were
corrected for demagnetization effects based upon crystal
dimensions. Our ac susceptibility measurements were found to be free
of Eddy currents effects as our measurements were independent of
excitation frequency and amplitude in the range of parameters
employed. The susceptibility of several crystals was measured in the
SQUID magnetometer with applied hydrostatic pressure, $P$, of up to
6.5 kbar in a beryllium-copper cell previously
described\cite{fecos2prb1}.

Shown in Fig.~1a is $\rho$ measured with the current in the ab plane,
$\rho_{ab}$, and along the the c-axis, $\rho_c$, of LaSb$_2$ as a
function of $T$ in zero magnetic field, $H$.  Note that the normal
state $\rho$ is highly anisotropic with $\rho_{ab}=1.2$ $\mu\Omega$ cm
at 4 K and $\rho_c / \rho_{ab} \sim 200$.  The $\rho_{ab}$ data
suggest a broad SC transition with an onset apparent near $T_{onset}
\sim 1.7$ K. However, a true $\rho=0$ state is not reached until ~0.7
K.  In contrast, the $T$ dependence of $\rho_c$ indicates an onset
near 1.0 K followed by a $\rho=0$ state below ~0.5 K.  Interestingly
the $\rho_c$ curve also shows a small peak for $T < T_{onset}$ similar
to was has been reported in (LaSe)1.14(NbSe2)\cite{szabo} and
attributed to a quasiparticle tunneling channel in the interlayer
transport. All of these features can be suppressed with the
application of magnetic fields as demonstrated in Fig.~1b where a
compelling difference in $\rho_{ab}$ and $\rho_c$ with $H$ oriented
along the ab planes is displayed. We observe that a field of $\sim
500$ Oe completely destroys the SC currents along c-axis while their
counterparts in the ab planes remain intact. This demonstrates a
relatively poor coupling between the SC condensate residing on
neighboring Sb planes.  We believe, in fact, that these data represent
two transitions: a planar superconducting transition initiating at
$T_{onset}$ and a secondary bulk transition below $T_c = 0.5$ K
associated with the emergence of coherent interlayer coupling.

\begin{figure}[htb]
  \includegraphics[angle=90,width=2.6in,bb=60 270 551
  687,clip]{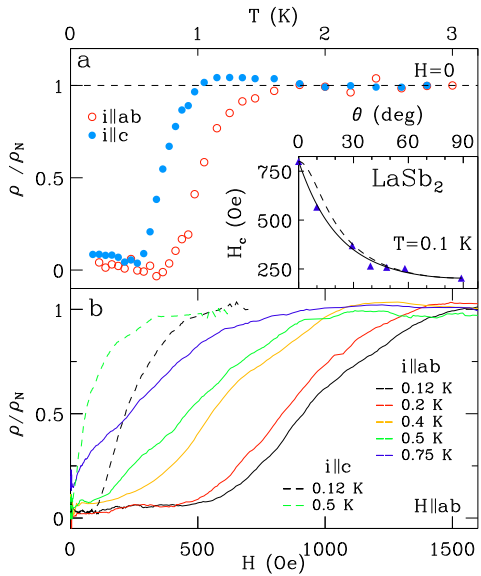}%
  \caption{\label{fig:resfig} Resistivity. (a) Resistivity, $\rho$,
divided by the normal state resistivity, $\rho_N$, vs temperature $T$
for currents along the ab-plane and the c-axis. Inset: Critical field,
$H_c$ vs angle, $\theta$, from $H$ parallel to the ab planes at 0.1
K. Solid (dashed) line is a plot of the 2D (anisotropic 3D) Tinkham
formula{\protect\cite{tinkham,supplmat}}. (b) $\rho / \rho_N$ vs
magnetic field, $H$, in the ab plane for currents perpendicular to $H$
in plane and along the c-axis. }
\end{figure}

\begin{figure}[htb]
  \includegraphics[angle=90,width=3.2in,bb=50 60 551
  737,clip]{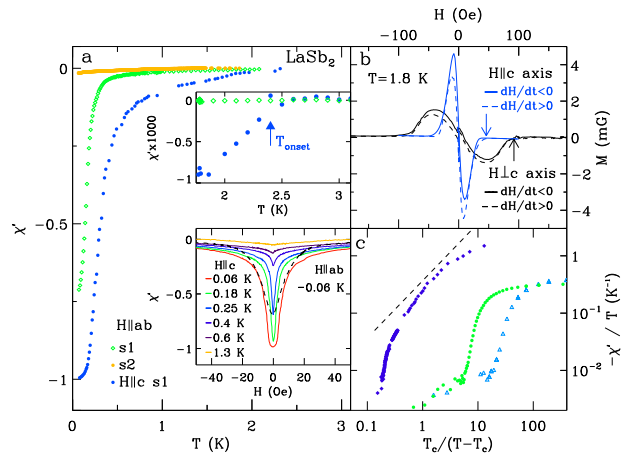}%
  \caption{\label{fig:tsweep} Ambient pressure susceptibility and
magnetization. (a) Real part of the ac susceptibility, $\chi'$, for
excitation fields along the c-axis and in the ab plane vs temperature,
$T$, for two representative crystals, s1 and
s2{\protect\cite{supplmat}}.  Upper inset: detail near the onset of
superconductivity, $T_{onset}$ as indicated by the arrow. Lower inset:
$\chi'$ for s1 vs magnetic field, $H$, at $T$'s identified in the
figure. (b) Magnetization, $M$, at $T=1.8$ K vs $H$ along the c-axis
and ab planes. (c) $-\chi'/T$ for $H \parallel$ c-axis vs reduced
temperature, $T_c / (T-T_c)$ for sample s1 with logarithmic axes at
$P=0$ (blue diamonds) and for a second sample with $P=2.7$ kbar (green
bullets), and $P=4.4$ kbar (blue triangles). Line is a linear
dependence. }
\end{figure}

Similar features are observed in the magnetic response of the SC phase
in LaSb$_2$, Fig.~2. However because $\chi$ and $M$ are more
representative of the true thermodynamic state of the system the
fragility of this phase results in a high sensitivity to growth
conditions, magnetic fields, and, as we show later, $P$.  Although all
crystals measured, more than 20, displayed $2.25 \le T_{onset} \le
2.5$ K in $\chi$ (Fig.~\ref{fig:tsweep}a upper inset), a broad range
of behavior was found in $\chi(T)$ with an incomplete Meissner effect
observed in most crystals. We believe that this sensitivity is due to
the proximity of LaSb$_2$ to a fully 3D SC phase. This disparate
behavior is demonstrated in Fig.~\ref{fig:tsweep}a where the real part
of the ac $\chi$, $\chi'$, is plotted for two of the 3 crystals whose
$\chi$ was explored at dilution refrigerator temperatures.  One
crystal, sample s1, displays a very broad transition to a $\chi'=-1$
state at $T< 0.2$ K for ac excitation fields, $H_{ac}$, oriented along
the c-axis. For $H_{ac}$ oriented along the $ab$ planes the
diamagnetic signal remains incomplete for s1, approaching $-0.75$ at
our lowest $T$, while the second sample, s2, displays only a small
diamagnetic signal. The full Meissner state in s1 for $H_{ac}
\parallel c$ is only apparent below 0.2 K despite a diamagnetism
consistent with type I superconductivity at $T< 2.5$ K as demonstrated
in Fig.~\ref{fig:tsweep}b. Here, similarly large anisotropies are
apparent in the magnetic field, $H$, dependence of $M$, that
faithfully reflect the crystalline structure.  The dc $H$ dependence
of $\chi'$ for s1 in the two field orientations are shown in the lower
inset to Fig.~\ref{fig:tsweep}a at a few $T$s. In
Figs.~\ref{fig:tsweep}a and b the small characteristic fields for the
destruction of the Meissner state are apparent.

Estimates based upon our previous $\rho(T,H)$, Hall
effect\cite{younglasb2}, and de Haas-van Alphen (dHvA)\cite{lasb2dhva}
measurements confirm our crystals have small carrier density, $n$,
small carrier mass, $m^*$, and highly metallic in-plane transport that
make anisotropic, type I, superconductivity sensible in LaSb$_2$. The
Hall coefficient with $H \parallel c$ is indicative of $n = 2\times
10^{20}$ cm$^{-3}$. The small $n$ and low $\rho_{ab}$ indicate highly
conductive transport along the ab plane with an estimated Hall
mobility of 2.7 m$^2$/Vs and mean free path, $\ell$, of $\sim 3.5$
$\mu$m\cite{younglasb2}. The reduction of the dHvA amplitudes with $T$
is small so that $m^*$ is only $0.2$ times the bare electron
mass\cite{lasb2dhva}. With these parameters, simple
estimates\cite{tinkham} of the London penetration depth, $\lambda$,
and Pippard coherence length, $\xi_0$, for currents in the ab plane
give $\lambda \ge 0.15 \mu$m, dependent on the SC condensate fraction,
and $\xi_0=1.6 \mu$m, much larger than in typical intermetallic
compounds. The large $\ell$ puts our crystals in the clean limit with
$\kappa = \lambda / \xi_0 <1$ consistent with type I superconductivity
and a small critical field, $H_c$. Type I superconductivity is rare in
intermetallic compounds and its discovery here is a reflection of the
extraordinarily long scattering times for currents in the ab
planes\cite{younglasb2,yonezawa}.

\begin{figure}[htb]
  \includegraphics[angle=90,width=2.75in,bb=80 230 535
  737,clip]{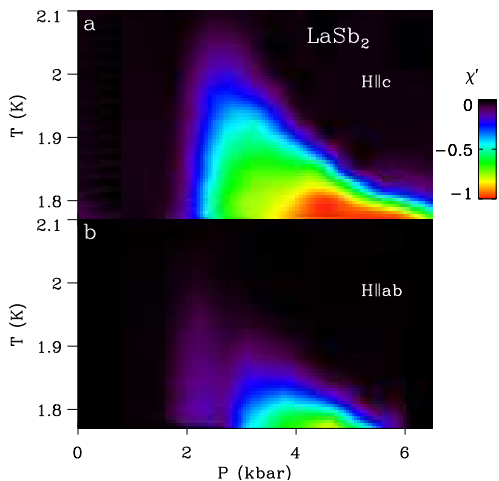}%
  \caption{\label{fig:press} Pressure and temperature dependence of
the superconducting transition.  Real part of the ac susceptibility,
$\chi'$, for magnetic fields $H$, along the c-axis (a) and along the
ab planes (b) vs pressure, $P$, and temperature, $T$. These contour plots
are produced by simple interpolation of measurements performed at
12 (11) different pressures in frame a (b). }
\end{figure}

The application of $P$ dramatically reduces the anisotropy and
significantly sharpens the transition as we demonstrate in
Fig.~\ref{fig:press}. Here we present the $P$ and $T$ dependence of
$\chi'$ for $T$s near the onset of superconductivity with the same
field orientations as in Fig.~\ref{fig:tsweep}.  Although we have only
followed $\chi'$ down to 1.78 K it is apparent that by 4.4 kbar the
transition width has been reduced to $\sim 0.1$ K with $\chi'=-1$ at
1.8 K for $H_{ac} \parallel c$, while for $H_{ac} \parallel ab$,
$\chi'<-0.75$. In addition, we do not observe the sample-to-sample
variability that was so apparent in the ambient pressure $\chi'(T)$.

The anisotropies that we measure in $\rho$ and $\chi'$ lead naturally
to the supposition that the SC in LaSb$_2$ is 2D. To address this
possibility we have determined $H_c$ by measuring $\rho(H)$ as a
function of field orientation at 0.1 K in the inset to 
Fig.~\ref{fig:resfig}a.  We observe a factor of 4 difference in $H_c$
as the crystal is rotated from an orientation where the ab planes are
nearly parallel to $H$ ($\theta=0$), $H_c^{\parallel}$, until they are
perpendicular to $H$ ($\theta=90 ^o$), $H_c^{\perp}$. For comparison
we plot the 2D Tinkham formula\cite{tinkham} prediction, solid line,
having no adjustable parameters beyond fixing $H_c^{\parallel}$ and
$H_c^{\perp}$ to match our data.  The sharp cusp in the data as
$\theta\rightarrow 0$ is a clear signature of 2D superconductivity. We
note that $H_c^{\parallel}$ is much smaller than the paramagnetic
limit which has been exceeded in some layered
materials\cite{prober}. Our measured $H_c^{\parallel}$ is likely
intrinsically limited by the long $\ell$ and large diffusion
constant\cite{tinkham}, and experimentally limited by the
flatness of our crystals.

\begin{figure}[htb]
  \includegraphics[angle=90,width=2.75in,bb=50 325 320
  695,clip]{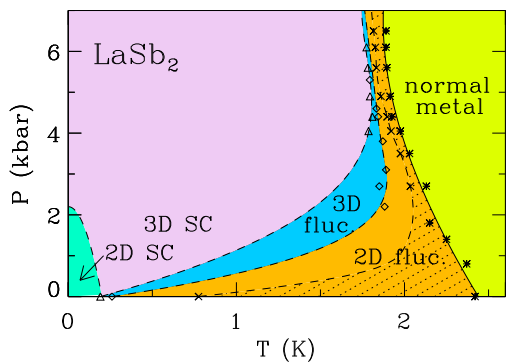}%
  \caption{\label{fig:pdiagb} Phase diagram. Proposed temperature,
$T$, and $P$, phase diagram. Symbols are onset of diamagnetism (*),
10\% (x's) and 90\% (triangles) of full Meissner for $H \parallel$
c-axis and 10\% of full Meissner $H \parallel$ ab planes
(diamonds). Lines are simple interpolations between the data points.
  }
\end{figure}

Our data, thus far, reveal LaSb$_2$ to possess an exceedingly unusual
SC phase characterized by large anisotropies for fields and currents
parallel and perpendicular to the Sb planes. The SC transition is
extraordinarily broad and, in the majority of samples, incomplete at
$P=0$ but is sharpened, and the anisotropy reduced, with application
of moderate $P$. In addition, the SC state at $P=0$ has an angular
dependent $H_c$ characteristic of a 2D superconductor along with
features in $\rho_c$ characteristic of quasiparticle tunneling between
Sb planes.  We believe that the interplane Josephson coupling of
essentially 2D SC planes mediates the high pressure 3D phase.  There
are other mechanisms for these observations that we have
considered. The first is the possibility that the SC state at $P=0$ is
restricted to the surfaces of the crystals and that a seemingly
unrelated 3D SC state is induced by the application of $P$. The large
Meissner fractions we observe in some of the samples and the
continuous evolution of the SC state with $P$ make this very
unlikely. Second, we have considered the possibility that we are
observing an anisotropic 3D SC state\cite{nbfilms} emanating from the
2D-like bands of LaSb$_2$\cite{lasb2dhva}. Anisotropic 3D
superconductivity is consistent with the ratio of
$H_c^{\parallel}/H_c^{c}$, but not the angular dependence in
Fig.~\ref{fig:pdiagb}a. In addition, it is difficult to explain the
large anisotropy in $\rho$ and $\chi'(T)$ in Figs.~\ref{fig:resfig}
and Fig.~\ref{fig:tsweep}a in in such a scenario. Finally, we point
out that the wide superconducting transition at ambient pressure is
not likely caused by impurities or second phases in our crystals since
our X-ray diffraction data are free from extraneous peaks, we deduce
very long mean free paths for carrier transport along the $ab$ planes,
and because the application of moderate pressure is unlikely to
suppress the effects of impurities or defects.

Thus, our data suggest that at low $T$ LaSb$_2$ is best described as a
set of Josephson coupled 2D planar superconductors. Interestingly, our
observation of an extraordinarily wide, and often times incomplete SC
transition at $P=0$, along with the dramatic changes apparent with
moderate $P$, indicate that the SC transition may be limited by phase
and amplitude fluctuations of the SC order parameter. Emery and
Kivelson have demonstrated that phase fluctuations are dominant when
the superfluid density is small so that there is small phase
stiffness\cite{emery}. Experiments have revealed that the underdoped
high $T_c$ SC cuprates are indeed phase fluctuation
limited\cite{orenstein}. The importance of phase fluctuations can be
determined by a comparison of $T_c$ with the $T=0$ phase stiffness,
$V_0 \propto L / \lambda^2$, which gives the $T$ at which phase
order would disappear, $T_{\theta}^{max}$.  Here, $L$ is the
characteristic length scale which in quasi-2D superconductors is the
larger of the spacing between SC layers or
$\sqrt{\pi}\xi_{\perp}$. With our estimated $\lambda$, and with the
assumption that $\xi_{\perp} < c/2=0.93$ nm, we find
$T_{\theta}^{max}\le 6.1$ times the onset $T$ for superconductivity at
ambient pressure (2.5 K), comparable to that tabulated for the
cuprates where $T_{\theta}^{max} / T_c$ ranges from 0.7 to
16\cite{emery}.

 One of the consequences of a phase limited transition is an extended
$T$-range where $\chi'$ is dominated by fluctuations at $T >
T_c$. Ginzburg-Landau (GL) theory, applicable in proximity to $T_c$,
predicts power-law dependencies for $\chi'/T$ in the reduced
temperature, $t = T_c /(T-T_c)$\cite{tinkham}. To check for such
power-laws in the $T$ range over which the SC phase develops we have
plotted $-\chi' / T$ as a function of $t$ for s1, where we have used
the maximum $\chi''(T)$ to define $T_c$, in
Fig.~\ref{fig:tsweep}c. The line in this figure represents the form
expected in 2D, $\chi'/T \propto t$. The data at ambient $P$ are well
described by a power-law form over a decade in $t$ with an exponent
that approaches that of the GL 2D prediction.  However, the large
values of $-\chi'$ that we measure, for example at $t\sim 1$ $-\chi'/T
\sim 0.1$, require $\xi_0\sim 11$ $\mu$m, about 7 times the estimate
based upon transport data. In contrast, the transitions at $P > 2$
kbar are not well described by a power-law in our range of $t$ as is
commonly the case when the SC state has a 3D character and the
fluctuation dominated regime is restricted to much larger $t$. The
phase diagram of Fig.~\ref{fig:pdiagb}b demonstrates how this picture
evolves with $T$ and $P$.

We conclude that at ambient pressure the very anisotropic SC phase of
LaSb$_2$ is fluctuation limited with fluctuations extending to $T$s an
order of magnitude greater than $T_c$. The small $m^*$, long $\ell$,
and small $n$ lead to large in-plane $\xi_0$ reducing the phase
stiffness of the SC state. The application of $P$ increases the
Josephson coupling between the SC planes leading to a more traditional
isotropic SC transition at the BCS $T_c$. Thus, our data suggest the
existence of a quantum, $T=0$, phase transition between 2D and 3D
superconducting phases with $P$. In addition LaSb$_2$ is a compelling
candidate for investigating the pseudogap region where SC pairs are
thought to form at $T$s above the phase ordering $T$, as in the
underdoped cuprates, in a BCS superconductor without the complication
of a competing ground state.

We are grateful to D. A. Browne and I. Vekhter for discussions. JFD,
DPY, and JYC acknowledge support from the NSF through DMR0804376,
DMR0449022, and DMR0756281. PWA acknowledges support from the DOE
through DE-FG02-07ER46420.


 \end{document}